\documentclass[aps,prl,twocolumn,superscriptaddress]{revtex4-2}
\usepackage{amssymb}
\usepackage{multirow}
\usepackage{bm}
\usepackage{amsmath}
\usepackage{graphicx}
\usepackage{epstopdf}
\usepackage{subfigure}
\usepackage{natbib}
\usepackage{epsfig}
\usepackage{amsfonts}
\usepackage{mathrsfs}
\usepackage[toc,page,title,titletoc,header]{appendix}
\usepackage[colorlinks,linkcolor=blue,citecolor=blue,anchorcolor=blue]{hyperref}

\usepackage{braket}

\usepackage{dsfont,amsthm,amsbsy}

\usepackage{fancyhdr}

\newcommand{\clbra}{ \Big\langle\!\!\Big\langle}
\newcommand{\clket}{ \Big\rangle\!\!\Big\rangle}

\begin{document}

\title{Models of interacting bosons with exact ground states: a unified approach}

\author{Zhaoyu~Han}
\affiliation{Department of Physics, Stanford University, Stanford, California 94305, USA} 

\author{Steven~A.~Kivelson} 
\affiliation{Department of Physics, Stanford University, Stanford, California 94305, USA} 

\begin{abstract}
We define an infinite class of ``frustration-free'' interacting lattice quantum Hamiltonians for bosons, constructed such that their exact ground states have a density distribution specified by the Boltzmann weight of a corresponding classical lattice gas problem. By appropriately choosing the classical weights, we obtain boson representations of various known solvable models, including quantum dimer and vertex models, toric code, and certain Levin-Wen string-net models. We also systematically construct solvable models with other interesting ground states, including quantum spin liquids, supersolids, ``Bose-Einstein insulators,'' Bose liquids with ``Bose surfaces'', and Bose-Einstein condensates that permit adiabatic evolution from a non-interacting limit to a Gutzwiller-projected limit.

\end{abstract}

\maketitle

Frustration-free Hamiltonians are an important class of solvable models whose ground states simultaneously minimize all the elementary components. In the modern study of quantum many-body physics, such models play an important role. For example, they were used to prove the existence and illuminate the physical properties of various strongly correlated phases of matter~\cite{Arovas1992}, such as Affleck-Kennedy-Lieb-Tasaki type models~\cite{PhysRevLett.59.799,PhysRevLett.60.531} of quantum disordered antiferromagnets, the Haldane pseudo-potential model~\cite{PhysRevLett.51.605} for fractional quantum Hall effects, the Rokshar-Kivelson quantum dimer model~\cite{PhysRevLett.61.2376,PhysRevB.35.8865} for resonating valence bond states, Kitaev models~\cite{KITAEV20062} and Levin-Wen string-net models~\cite{PhysRevB.71.045110,PhysRevB.103.195155} for topological orders, Haah's code~\cite{PhysRevA.83.042330} and X-cube model~\cite{PhysRevB.94.235157} for fractons, ``quantum geometric nesting'' models~\cite{PhysRevX.14.041004} for fermion bilinear orders in flat-band systems, among many others~\cite{ARDONNE2004493,PhysRevB.43.3255,PhysRevB.76.134514,CASTELNOVO2005316,PhysRevB.84.235145,PhysRevLett.105.060504,PhysRevLett.125.176402,Zhang_2023,mielke1993ferromagnetism, PhysRevB.92.035115, PhysRevB.94.115127,PhysRevB.108.075142,PhysRevB.109.085106,PhysRevLett.108.247206,PhysRevLett.89.137202,han2023fractional,PhysRevB.105.155130,PhysRevResearch.3.033190,PhysRevB.88.165303,PhysRevB.88.115133,PhysRevB.89.195130,10.21468/SciPostPhysCore.4.4.027,PhysRevB.84.165139,PhysRevB.94.245149,PhysRevB.96.115107,PhysRevB.92.155105,PhysRevB.90.245122,watanabe2023spontaneous}. Frustration-free points also appear in the parameter space of certain strongly correlated problems, providing convenient handles for investigations of nearby regimes~\cite{PhysRevB.107.125134,ran2024hidden,PhysRevB.109.L241109,yan2022triangular,PhysRevB.106.L041115,PhysRevLett.91.167004,PhysRevB.69.224415,PhysRevB.69.224416,doi:10.1073/pnas.2015785118,PhysRevX.11.031005,PhysRevX.12.041029,PhysRevLett.122.246401,herzog2022many,PhysRevX.10.031034,PhysRevLett.131.106801,PhysRevLett.130.186404,han2024emergent}. For example, the prediction of a quantum spin liquid in a Rydberg array~\cite{doi:10.1126/science.abi8794} was made using this sort of methodology~\cite{PhysRevX.11.031005,doi:10.1073/pnas.2015785118,PhysRevX.12.041029}. In recent years, the special entanglement structure of such models~\cite{PhysRevA.84.042338,10.1063/1.4748527,10.1145/3519935.3519962,10.1063/1.4922508} has also drawn attention to their possible applications to quantum information processing~\cite{Darmawan_2012,Darmawan_2014,Thibodeau2023nearlyfrustration,Zhu2024efficient,10.1145/3357713.3384292}. 

However, frustration-free models are hard to find and even to diagnose (known as the quantum satisfiability problem~\cite{doi:10.1073/pnas.1519833113,bravyi2006efficient}). Systematic construction schemes of such models are thus desirable. 
In this work, we construct an infinite class of frustration-free, interacting lattice models for bosons by (roughly speaking) relating the problem to Bose-Einstein condensation (BEC) of free bosons, adopting a method spiritually similar to Witten's conjugation argument~\cite{10.21468/SciPostPhysCore.4.4.027,WITTEN1982253} for generalizing supersymmetric models. The models can be relatively simple and physical, consisting of short-range boson hopping terms and few-body density interactions. The resulting ground state is a BEC state with a re-weighted density distribution identical to a classical lattice gas problem. Most models constructed in this way are gapless with an anomalously soft density mode, but they can also be gapped when there are hard constraints in the problem. We will show that this scheme unifies many known frustration-free models, leads to their natural generalizations, and furthermore allows constructions for other interesting phases of matter. Necessary technical details can be found in the End Matter.

% Below, we first present the general discussions of the formalism of the construction scheme and properties of the constructed models and then discuss how it can be applied to specific states of matter. 

{\bf The models: } The models we consider can be defined on {\it any} lattice in {\it any} dimension. In the present work, we will consider systems that conserve boson number, but the scheme can be readily generalized to systems of bosons whose number is not conserved (e.g., spin waves, phonons, or photons), as discussed in App.~\ref{app: number non conserving}. The models are defined by Hamiltonians of the form:
\begin{align}
    \hat H =& \sum_{c,c'}\tau_{c,c'} \hat{H}_{c,c'}\\
    \hat{H}_{c,c'} \equiv& \left[-\hat{T}_{c,c'} + \hat{V}_{c,c'}\right]
\end{align}
where the sum runs over pairs of equal-size clusters of sites $c$ and $c'$, $\hat{T}$ is a ``kinetic energy'' term that hops bosons between clusters $c$ and $c'$, $\tau_{c,c'}$ is an {\it arbitrary} non-negative amplitude (which we will take to be finite-ranged), and $\hat{V}_{c,c'}$ is the corresponding ``potential energy,''  i.e. it is diagonal in the boson occupation configuration basis $|\{n\}\rangle$. We identify clusters with $L$-tuples of site indices, $c = (i_1\dots i_L) $ (where a site index might occur multiple times in this unordered
list). In the simplest cases, the clusters consist of individual sites, $c,c' = i,j$, such that $\hat T_{i,j} = \left(\hat b_i^\dagger \hat b_j + \text{h.c.}\right)$, where $\hat b_i^\dagger$ creates a boson on site $i$. % and $\hat n_i=\hat b_i^\dagger \hat b_i$ is the number density operator. 
The key bit of reverse engineering concerns the choice of the operator $\hat{V}_{c,c'}$ for which a certain interesting state is the exact ground state.

Specifically, the solvable Hamiltonian is designed by working backward from an {\it arbitrary real} function, $K(\{n\})$, of the boson occupation configuration $\{n\}$. Ultimately, we will find that  $\mathrm{e}^{-K(\{n\})}$ is the Boltzmann weight of an associated classical lattice gas problem defined on the same spatial lattice, which has the same density correlations as the ground state of the quantum problem. The kinetic terms associated with the pair of clusters $c$ and $c'$ is 
\begin{align}
     &\hat{T}_{c,c'} =  \hat{B}^\dagger_{c}  \hat{B}_{c'}  +  (c \leftrightarrow c') \nonumber\\
    &\hat{B}_{c}
    \equiv  \prod_{k\in c} \hat{b}_{k}
\end{align}
i.e. it hops $L$ bosons from $c$ to $c'$ or conversely. The interaction term is
\begin{align}
    \hat{V}_{c,c'} 
    =  \hat{B}^\dagger_{c} \ \mathrm{e}^{
 \left[\delta K_{c}(\{\hat{n}\}) - \delta K_{c'}(\{\hat{n}\})\right]/2} \ \hat{B}_{c} +  (c \leftrightarrow c')
\end{align}
where $\delta K_c$ is the change of $K$ produced by the change in the configuration corresponding to the operation of $B^\dagger_c$,
\begin{align}
  \delta K_c(\{n\}) \equiv K(\{n\}_c) - K(\{n\})
\end{align}
where $\{n\}_c$ is defined as the occupation configuration of $\hat{B}_c^\dagger|\{n\}\rangle$. 

To make the structure of the Hamiltonian clear, it is convenient to define a set of `soft projection' operators (which are non-negative, and mutually commuting)
\begin{align}
    \hat{P} \equiv \mathrm{e}^{-K(\{\hat{n}\})/2}\ \ \ \&\ \ \
    \hat{P}_c \equiv  \mathrm{e}^{-\delta K_c(\{\hat{n}\})/2}.
\end{align}
Then, using the equality $\hat{P}_c  \hat{B}_c = \hat{P}^{-1} \hat{B}_c \hat{P}$, the Hamiltonian can be conveniently recast into a form that is reminiscent of Witten's conjugation construction~\cite{10.21468/SciPostPhysCore.4.4.027,WITTEN1982253} (for a formal discussion, see App.~\ref{app: conjugation})
\begin{align} \label{eq: recasted Hamiltonian}
    \hat{H}_{c,c'} = & \hat{\mathcal{B}}^\dagger_{c,c'} \hat{\mathcal{B}}_{c,c'} \\
    \hat{\mathcal{B}}_{c,c'} \equiv & \left(\hat{P}_{c}\hat{P}_{c'}\right)^{1/2} \hat{P}\left(\hat{B}_{c} - \hat{B}_{c'}\right)\hat{P}^{-1}\label{eq: semidefinite}
\end{align}
so they are always positive semi-definite. 

Note also that there is no need, anywhere in the above construction, to stick with clusters with a given character or number of operators, $L$. It is perfectly possible for $\hat{H}$ to simultaneously contain different types of terms hopping different numbers of bosons. Moreover, if both $\tau$ and $K$ are short-ranged, then, of course, so is $\hat H$.

{\bf Ground states: } The ground states of the Hamiltonian are a set of coherent states modified by $\hat{P}$
\begin{align}\label{eq: coherent state}
    \left|\alpha \right\rangle \equiv & \frac{1}{\mathcal{N}_\alpha}\ \hat{P} \exp\left[\alpha \sum_{i}\hat{b}^\dagger_i \right] \left|0\right\rangle 
    \ . 
\end{align}
where $\alpha$ is an arbitrary complex number, $\mathcal{N}_\alpha$ is a normalization factor. To see why these states are ground states and the models are frustration-free, we note that each $ \hat{H}_{c,c'} $ term in the Hamiltonian ends with $\left(\hat{B}_{c} - \hat{B}_{c'}\right)\hat{P}^{-1}$ that annihilates $|\alpha\rangle$
\begin{align}
   &\left(\hat{B}_{c} - \hat{B}_{c'}\right) \hat{P}^{-1} |\alpha\rangle \nonumber \\
= &\frac{1}{\mathcal{N}_\alpha} (\alpha^L-\alpha^L) \exp\left[\alpha \sum_{i}\hat{b}^\dagger_i \right] \left|0\right\rangle =0 .
\end{align} 
It thus follows that $|\alpha\rangle$, which is a modified BEC state, is a ground state of $\hat{H}$, since each $\hat{H}_{c,c'}$ is non-negative and $\hat{H}_{c,c'}  |\alpha\rangle = 0 $. Since the system conserves boson number, one can also construct a ground state within each number sector. We discuss this construction and its properties in App.~\ref{app: number basis}.

As shown in the App.~\ref{app: mapping to classical}, the normalization factors of these wavefunctions are related to the partition function of a (generalized) classical lattice gas problem:  $\mathcal{N}^2_\alpha$ is equal to the grand-canonical partition function, 
\begin{align}
   \mathcal{N}^2_\alpha    =
   \mathcal{Z}_\alpha \equiv 
   \int {\cal D}\{n\}\ 
   \mathrm{e}^{- K(\{n\}) + \mu_\text{cl}     \sum_{k} n_k}  
\end{align}
with `chemical potential' $\mu_\text{cl} \equiv \ln(|\alpha|^2)$ and where the sum over configurations of the lattice gas of indistinguishable particles, $\int {\cal D}\{n\}\equiv \sum_{\{n\}} \left[\prod_k\frac{1}{ n_k!}\right] $, includes a statistical factor in the measure~\footnote{The factorial factors are present because we are summing over occupation configurations, not the particle position configurations.}.

It is interesting to note that the ground-states are determined by $K(\{n\})$ and independent of $\tau_{c,c'}$, which means that the ground states can have higher symmetry than the Hamiltonian. As a dramatic example, one can construct a disordered Hamiltonian with spatially random hopping matrix elements, $\tau_{c,c'}$, while keeping $K(\{n\})$ translationally invariant, which results in a uniform ground state (similar to the Mattis model~\cite{MATTIS1976421}).

{\bf Correlation functions: } All the density correlations in these ground states are identical to those in the corresponding classical problem. Specifically, 
\begin{align}
   \Big\langle C(\{\hat{n}\})\Big\rangle = & 
   \clbra C(\{n\}) \clket 
\end{align}
where $C(\{n\})$ is an arbitrary function of occupation configurations, and 
$\clbra \dots \clket $ represents the thermal average in the classical ensemble defined by $K$. Therefore, whenever the classical system has certain spontaneous symmetry breaking by density modulations or certain constraints on the density distribution, so do the corresponding quantum ground states.

The boson correlation functions can also be explicitly related to certain density correlation functions in the corresponding classical problem. For example, for clusters $c$ and $c'$ of size $L$ (see the SM for more general expressions):
\begin{align}
G_{c,c'}\equiv &
      \left\langle \alpha\left| \hat{B}^\dagger_{c} \hat{B}_{c'}    \right|\alpha\right\rangle = \mathrm{e}^{L \mu_{\text{cl}}} 
    \clbra 
    P_c(\{n\}) P_{c'}(\{n\})\clket .
\end{align}

When $K$ is short-ranged, barring the taking of a singular limit in which some parameter in $K$ tends to infinity (i.e. enforces a constraint), one generally expects that the connected correlators of the classical problem decay with distance,  meaning that 
\begin{align}\label{eq: ODLRO}
    G_{c,c'} \sim \clbra P_c(\{n\}) \clket ~  \clbra P_{c'}(\{n\}) \clket
\end{align}
for far-separated clusters - i.e., the ground states in question almost always exhibit off-diagonal long-range order (ODLRO), so these are generalized BEC states.

{\bf Dynamical density fluctuations:} Besides the ground states, the dynamics of the density fluctuations can also be inferred. For concreteness, in this and the next section, we will consider the case of inversion symmetric systems on simple periodic lattices so that a single Bloch wave vector, $\bm{q}$, can label the density mode fluctuation. The density operator at momentum $\bm{q}$,
\begin{align}
    \hat{\rho}_{\bm{q}} \equiv \sum_i \hat{n}_i \mathrm{e}^{-\mathrm{i} \bm{q}\cdot \bm{r}_i} 
\end{align}
($\bm{r}_i$ is the position of lattice site $i$), applied
to any ground state $|\alpha\rangle$ with density $\bar{n}$, yields a state $\hat{\rho}_{\bm{q}}|\alpha\rangle$ with momentum $\bm{q}$. The variational energy of this state provides a rigorous upper bound on the lowest energy in this momentum sector, which can be evaluated as (see App.~\ref{app: dynamical density modes} for a detailed discussion):
\begin{align}
    E_\text{SMA}(\bm{q})  = \frac{f(\bm{q})}{s(\bm{q})}
\end{align}
where ($g_c(\bm{q}) \equiv \sum_{j\in c}\mathrm{e}^{-\mathrm{i} \bm{q}\cdot \bm{r}_j}$, $\mathsf{V}$ is the number of sites)
\begin{align}
    s(\bm{q}) =& \frac{1}{\mathsf{V}} \sum_{ij}  \clbra n_i n_j \clket \mathrm{e}^{\mathrm{i} \bm{q}\cdot (\bm{r}_i-\bm{r}_j)} \\
f(\bm{q}) = & \frac{1}{2\mathsf{V}} 
 \sum_{c,c'} \tau_{c,c'}\left| g_c(\bm{q})-g_{c'}(\bm{q}) \right|^2 
\left(G_{c,c'}+ \text{h.c.} \right) \label{eq: fofq}
\end{align}
are the equal-time structure factor and the first frequency moment of the dynamical structure factor, respectively. $E_\text{SMA}(\bm{q})$ is readily calculable from a solution (if needed, a numerical solution) of the classical problem. The famous single-mode approximation (SMA) amounts to adopting this variational energy as the actual excitation energy~\cite{Arovas1992,feynman2018statistical,girvin1986magneto}.  The accuracy of this approximation can, in principle, be systematically estimated by calculating higher frequency moments of the dynamical structure factor, which can also be expressed in terms of correlation functions of the related classical problem. 

In the limit $\bm{q}\rightarrow \bm{0}$, for generic forms of short-range $\tau$ and $K$, it follows that $f(\bm{q}) \sim \bm{q}^{2}$ and $s(\bm{q}) \rightarrow s_0$ where $s_0\equiv \frac{\partial \bar{n}}{\partial \mu}$ is the compressibility of the classical systems. Thus, so long as the classical system is compressible, the density excitation spectrum of the quantum system at $\bm{q}\rightarrow \bm{0}$ must vanish at least as fast as $\bm{q}^2$~\footnote{This is consistent with several proven bounds for frustration-free systems~\cite{PhysRevLett.116.097202,masaoka2024rigorous}. }
 
Naturally, if the classical system develops CDW long-range order with an ordering vector $\bm{Q}$ it follows that $E_\text{SMA}(\bm{q}) = E_\text{SMA}(\bm{q}+\bm{Q})$.  More interestingly, if the classical thermal state is translationally invariant but has a long CDW correlation length (i.e. is nearly critical), it follows that $s(\bm{q})$ is strongly peaked at $\bm{q}=\bm{Q}$, implying a corresponding ``roton-minimum'' in $E_\text{SMA}(\bm{q})$.

{\bf Stability analysis:}
When viewing these solvable models as special submanifolds in the vast parameter space of interacting boson problems, it is natural to ask how deviations from these solvable points affect the physics.  This can be addressed using a mean-field treatment of the perturbations within the SMA framework~\cite{Arovas1992}. Alternatively, here we analyze the effective {\it quantum} field theory of the density fluctuation field $\delta n \equiv n-\bar{n}$ and its canonical conjugate - the boson phase field $\theta$. Under the assumption that a quadratic action suffices to describe the low-energy physics, we obtain the following effective theory (expressed in Euclidean spacetime) that reproduces the $s(\bm{q})$ and $f(\bm{q})$ factors already discussed:
\begin{align}
    S_\text{eff}
    = \sum_{q \equiv (\rho_n,\bm{q})} \omega_n \delta n_{q} \theta_q^* 
    + \frac{f(\bm{q})}{2} \left[\frac{\left|\delta n_{q}\right|^2}{s^2(\bm{q})} +  \left| \theta_q\right|^2\right]\ .
\end{align}

For compressible classical systems, $s_0 $ is finite while $f(\bm{q}\rightarrow 0)\sim  \kappa \bm{q}^2$,  so the long-wavelength limit of the theory is
\begin{align} \label{eq: seff}
    S_\text{eff}    \sim \int \mathrm{d}\tau\mathrm{d}\bm{r} \left\{\mathrm{i} \delta n \cdot \dot{\theta} + \frac{\kappa}{2}\left[\frac{\left(\nabla\delta n\right)^2}{s_0^2} + \left(\nabla \theta\right)^2\right]\right\}
\end{align}
which is the same as that of free bosons, has an enlarged ISO(2) symmetry of the charge U(1)~\cite{sym2020609}, and describes an unstable fixed point from the renormalization group perspective. However, the only allowed relevant perturbation is $w(\delta n)^2/2$. (A $\theta^2$ term is relevant but not allowed since it violates charge conservation.) A positive $w$ stabilizes a normal superfluid phase with superfluid stiffness $\kappa$ and a linear sound mode dispersion with velocity $v=
\sqrt{w\kappa}$. However, a negative $w$ makes the system unstable to phase separation. We thus conclude that the solvable models generically describe a system on the boundary of a superfluid phase.

Although it is possible, it is much more subtle to write down the correct effective quantum field theory when the classical system is incompressible, which can happen under appropriate circumstances when there are hard constraints (i.e. the classical weight $K(\{n\})$ contains infinite terms) or in the presence of long-range interactions. At least in some cases, this admits to a description in terms of emergent gauge fields and brings with it associated notions of fractionalization, confinement-deconfinement transitions, and topological order. For previously solved examples, we refer readers to the literature on quantum spin liquids and ices~\cite{moessner2010quantum, fradkin2013field,savary2016quantum, han2024emergent,balasubramanian2022exact,PhysRevB.65.024504,doi:10.1142/S0217984990000295}.

{\bf  Applications:}  We now apply this general approach in a variety of interesting cases.

\underline{Superfluid with short-range repulsions:} An obvious application is to bosons with strong short-range repulsions.  A widely used class of variational states that have been applied to such problems involves multiplying a simple BEC state by an appropriate Jastrow factor that builds in the requisite short-range correlations.  For bosons living on the vertices of a regular lattice, the simplest such Jastrow factor corresponds to $\hat{P}$ defined by
\begin{align}
    K(\{n\}) = u \sum_{i} (n_i-\bar n)^2
\end{align}
which suppresses the bunching of bosons on a given site. The corresponding reverse-engineered potential for a pair of single-site clusters (e.g. nearest neighbors) is 
\begin{align}
    \hat{V}_{i,j}=    \hat n_i e^{u(\hat n_i-\hat n_j - 1)}  +\left(i\leftrightarrow j\right)
\end{align}

There are two interesting implications of this:  1) The first is that it provides an example of a system of strongly interacting bosons with ground-state properties that are still - in effect - equivalent to those of a non-interacting BEC.  As already mentioned, the way to interpret this is as plotting a fine-tuned trajectory through Hamiltonian space lying precisely on the boundary of the superfluid phase - where almost any form of additional attractive interactions will trigger an instability to phase separation.  2) In the limit  $u\rightarrow \infty$, the corresponding $\hat{P}$ operator becomes a Gutzwiller projection that constrains $n_i$ on each site to be either $p$ or $p + 1$, where $p=\lfloor \bar{n} \rfloor$ is the largest integer that is no larger than the average particle number per site $\bar{n}$. This establishes the possibility of adiabatic evolution from the weakly interacting limit to a hard-core limit of the BEC ground state, despite arguments that the truncation of the Hilbert space (i.e. a change in the ``exclusion statistics''~\cite{PhysRevLett.67.937}) that characterizes this limit must fundamentally change the emergent properties.  An obvious exception in which the $u\to \infty$ limit does indeed change the character of the state arises in the case that $\bar n$ is an integer, in which case the classical system becomes incompressible in this limit, corresponding to a Bose-Mott insulating phase.
\footnote{For $\bar{n}$ an integer and $u$ large but non-infinite, the system still possesses ODLRO, but with $\langle b_k\rangle \sim e^{-u}$ - i.e. it is a form of ``gossamer superfluidity.''~\cite{laughlin2006gossamer}}

\underline{Supersolids:}
This approach also naturally constructs models for supersolids -- states with coexisting CDW and superfluid order. For example, this is achieved on a regular lattice with
\begin{align}\label{eq: K for supersolid}
    K(\{n\}) = u \sum_{i} (n_i-\bar n)^2 + v\sum_{\langle ij\rangle } (n_i-\bar n)( n_j-\bar n)
\end{align}
which, interpreted as a classical lattice gas Hamiltonian, clearly gives rise to a CDW phase for large enough $v>0$. 
In particular, in the $u\rightarrow \infty$ limit, the classical lattice gas problem maps to the Ising model (with the two possible occupations on each site, $n_i =p$ or $p + 1$), which has been investigated in depth on various lattices and is known to have crystalline phases, both with commensurate and incommensurate wavevectors~\cite{PhysRevB.28.2897}. In this case, the reverse-engineered potential for a pair of nearest-neighboring sites is 
\begin{align}
    \hat{V}_{i,j} = &  \hat{n}_i  \mathrm{e}^{(\infty)(\hat n_i-\hat n_j - 1)+v (\hat{N}_i-\hat{N}_j+1)/2}+(i\leftrightarrow j) 
\end{align}
where $\hat{N}_i \equiv \sum_{k \in \text{ nn } i} \hat{n}_k$ is the total particle number on the neighboring sites of $i$. [Since all the $\hat{n}_i = p,  p +1$, $\hat V$ can be expressed in a polynomial form.] 

On a bipartite lattice with $\bar n=1/2$, the CDW expected for large $v$ is commensurate, corresponding to one boson per emergent unit cell.  This again evolves into a Bose-Mott insulating state as $v\to \infty$, but for any finite $v$, it still exhibits ODLRO.

\underline{Quantum dimer models :} 
Gutzwiller projection is not the only case in which the present approach can be used to study model problems defined in constrained Hilbert places. Here, we show how this scheme can also unify quantum dimer models (QDM). In this case, the sites are taken to lie on the edges of a given lattice (forming its generalized `Lieb lattice'), and the bosons are considered `dimers.' The dimer covering constraint can be achieved by 
\begin{align} \label{dimer}
    K(\{n\}) = &u \sum_{i} \left( n_i-\bar{n}\right)^2 + u' \sum_{\mathsf{v}} \left(N_\mathsf{v}-\bar{N}\right)^2
\end{align}
where $N_\mathsf{v} \equiv   \sum_{i \in \partial \mathsf{v} } n_i $ is the number of dimers occupying the edges connected to the vertex $\mathsf{v}$ (represented by $i \in \partial \mathsf{v}$), and $\bar{N} =\bar{n} \mathsf{C}/2 $ is the average number of $N_\mathsf{v}$ ($\mathsf{C}$ is the coordination number of the lattice). For $0< \bar{n} <1$, the corresponding $\hat{P}$ projects onto states with at most one dimer on each edge as $u\rightarrow \infty$, and either $\lfloor \bar{N}\rfloor$ or $\lfloor \bar{N}\rfloor+1$ dimers touching each vertex as $u'\to \infty$.  Specifically, for $\bar{N}=1$, a hard-core dimer constraint is enforced, and hence the reference classical statistical mechanics model is the much-studied hard-core dimer model.  On the other hand, for $\bar{N}=2$, the corresponding classical model sums over non-intersecting, undirected, fully-packed loops, another famous model in classical statistical mechanics~\footnote{More generally, for any $\bar N \leq 2$, the reference classical statistical mechanics model corresponding to undirected, non-intersecting loops at general packing can be obtained by substituting $\left(N_\mathsf{v} -\bar N\right)^2 \to N_\mathsf{v}\left(N_\mathsf{v}-2\right)^2$ in Eq.~\ref{dimer}, and then taking the familiar limit $u$ and $u'\to \infty$.}.

Any Hamiltonian constructed in this way is effectively tuned to a generalized ``Rokhsar-Kivelson point''~\cite{PhysRevLett.61.2376,PhysRevB.35.8865}, i.e. one for which the ground state is an equal superposition of all projected (constraint-satisfying) states. One subtlety arises from the fact that for integer values of $\bar{N}$, single-boson hopping terms do not operate within the constrained Hilbert space, so multi-boson hopping terms, such as dimer ``resonances'' around plaquettes (which are ``dipole conserving''), must be included.

\underline{Other models with constrained Hilbert spaces:}  Further extending this methodology, one can also obtain boson representations of certain Levin-Wen string-net models~\cite{PhysRevB.71.045110,PhysRevB.103.195155}, Kitaev's toric code~\cite{KITAEV20062} or quantum vertex models~\cite{PhysRevB.106.195127,ARDONNE2004493,PhysRevB.69.064404,balasubramanian2022exact}. The strategy is to introduce several flavors of bosons on the lattice edge to represent the possible `string types' and then use an appropriate $K(\{n\})$ to impose a single-occupancy constraint on each edge and a constraint at each vertex so that surrounding string types obey the `branching rules.' By considering the quantum resonance of the strings consistent with the rules, one can, in this way, naturally obtain certain solvable Hamiltonians for deconfined gauge theories, whose groundstates have trivial relative phases between different string configurations.

\underline{Dipole conserving dynamics:}  It is sometimes interesting to consider Hamiltonians with dynamics that preserve the center of mass position.  Models of this sort can arise in the low energy effective theory of strongly interacting particles where remaining dynamical processes must satisfy strong constraints.  Examples of this are the ring-exchange processes in quantum crystals~\cite{Thouless_1965},  the ``resonance'' terms in the QDM~\cite{PhysRevB.35.8865}, induced pair-hopping terms in quantum Hall systems in the thin cylinder limit~\cite{Bergholtz_2006,PhysRevB.77.155308,PhysRevLett.95.266405}, and the hopping processes in a strongly tilted optical lattice~\cite{PhysRevB.101.174204}.

Many aspects of systems with this sort of kinetic energy can be treated from the present perspective by suitably defining the clusters $c$ and $c'$ connected by $\hat{H}_{c,c'}$. The resulting solvable ground states can still possess ODLRO. However, for processes that conserve the center-of-mass position of the bosons, 
$g_c(\bm{q}) - g_{c'}(\bm{q})$ vanishes at least as $\bm{q}^2$ as $\bm{q} \to \bm{0}$, implying that $f(\bm{q}) \sim |\bm{q}|^4$ as $\bm{q}\to \bm{0}$ (see Eq.~\ref{eq: fofq}). It follows from Eq.~\ref{eq: seff} that the superfluid stiffness (i.e. the coefficient of $\left({\bf \nabla} \theta\right)^2$  in the effective action)  vanishes. Since any uniform current-carrying state would violate the assumed dipole conservation, such states are also, in some sense, insulating - they have thus been named `Bose-Einstein insulators~\cite{PhysRevB.106.064511}'.

\underline{Bose liquids with `Bose surfaces':} When the non-zero $\tau_{c,c'}$'s all correspond to $\left|g_c(\bm{q}) - g_{c'}(\bm{q})\right|^2$ that vanish on a submanifold $\mathcal{M}$ of the momentum space, this implies that $f(\bm{q})$ as well as $E_\text{SMA}(\bm{q})$ (assuming the classical system is compressible) vanish on $\mathcal{M}$. For most cases, $\mathcal{M}$ is a single point at $\bm{q}=\bm{0}$, but sometimes it can be higher dimensional. As a concrete example, consider short-range resonant pair hopping on a square lattice where a pair of bosons from one set of opposite vertices, $c$, of an elementary plaquette is shifted to the complementary vertices, $c'$. [Note that this process happens to be dipole conserving; moreover, if we identify the bosons with dimers and the lattice vertices with the edge centers of another rotated square lattice, this is the sort of dimer resonance term mentioned above that must be introduced to obtain the dynamical terms in the QDM at $\bar N=1$.] The resulting $f(\bm{q}) \sim \sin^2 q_x \sin^2 q_y$ not only vanishes as   $\bm{q}\rightarrow \bm{0}$, but also along lines in $\bm{q}$-space. Therefore, the system has gapless modes along one-dimensional `Bose-surfaces'~\cite{PhysRevB.75.235116} in the Brillouin zone.

We now analyze the stability of this behavior:

In the context of the above example on the square lattice, $\mathcal{M}$ is robust against the inclusion of resonant tunneling processes that shift bosons along nearest-neighbor bonds around larger loops.  However, other types of boson-hopping processes generically induce fundamental changes to $\mathcal{M}$. For example, adding a usual single-particle boson hopping term, or adding resonant pair-tunneling involving bosons moving to second-nearest-neighbor sites around a suitably defined loop (bounding a $\sqrt{2} \times \sqrt{2}$ plaquette), will reduce $\mathcal{M}$ to one (or more) isolated point.

The addition of a density interaction $w (\delta n)^2/2$ to the effective action makes the system unstable to phase separation (as usual) for attractive $w<0$. In contrast, for repulsive $w>0$, it keeps the gapless manifold $\mathcal{M}$ but changes the boson excitation dispersion near $\mathcal{M}$ from quadratic to linear: $E(\bm{q}) \sim \sqrt{wf(\bm{q})}$, which results in a spatially decaying boson Green's function $\propto \langle \mathrm{e}^{\mathrm{i}[\theta(\bm{r})-\theta(\bm{0})]}\rangle$ with a similar character as for a Fermi liquid with the same manifold  $\mathcal{M}$ of gapless modes (e.g. a surface in 3D or a nodal line in 2D)~\footnote{It should be noted that the state constructed here is different from the Bose liquid studied in Ref.~\cite{PhysRevB.75.235116} in many aspects, including its specific heat and the relation between the particle density and the volume enclosed by the Bose surfaces. }.

%In this work, we have developed a unified approach for constructing ``frustration-free'' Hamiltonians for number-conserving bosons whose exact ground-state properties can be inferred from the properties of an associated classical statistical mechanical problem on the same lattice.

{\bf {Discussion:}} For the known solvable models, our scheme provides not only an alternative representation that might be realizable on engineered platforms like optical lattices but also a natural route from a strongly constrained problem to a free (trivial) limit of the boson problem. When the constrained limit of the problem exhibits an exotic quantum phase of matter, this evolution must drive exotic transitions out of them, whose properties may be studied with these engineered models. The critical points constructed this way are ``quantum Lifshitz points''~\cite{ARDONNE2004493} with unusual dynamical properties~\cite{CASTELNOVO2005316,PhysRevB.83.125114}. 

The asymptotic behaviors of correlations in $d$ dimensional quantum ($T=0$) and classical ($T> 0$) systems are generically different.  Thus, solvable models of the sort we have discussed are always, to some degree, non-generic. Indeed, in many cases, as we have shown, the solvable models describe systems on the phase boundary beyond which there is an instability - often to phase separation. A dramatic manifestation of this point is that the charge $U(1)$ symmetry is broken in the ground state of such models even in one dimension, similar to certain frustration-free spin models~\cite{PhysRevLett.133.176001} and ``transverse quantum fluids''~\cite{PhysRevB.109.214519}. Still, it is possible to infer generic properties of interesting phases by constructing an effective field theoretical representation of the solvable model and then addressing the effect of symmetry-allowed perturbations.

A limitation of the specific construction in the current work is that the relative phase between different configurations is trivial, which can be overcome by considering complex-valued $K(\{n\})$ functions and/or starting from other frustration-free boson hopping models with zero mode(s) with non-trivial phase structures. We will leave exploration of these generalizations to future work.

% We have not, as of yet, developed a similar systematic approach to generating fermionic models, or even bosonic models with kinetic frustration (e.g. in which the bosons experience an orbital magnetic field) whose ground-states can be related to simple classical Boltzmann weights.  That such a generalization is possible seems clear;  the density-density correlations of the Laughlin state (bosonic or fermionic) for the fractional quantum Hall effect are equal to those of a corresponding classical one-component plasma~\cite{PhysRevLett.50.1395}. Moreover, engineered models for which the Laughlin state is exact have been constructed in a variety of ways~\cite{PhysRevLett.51.605,PhysRevB.31.5280}. 

{\bf Acknowledgements:}  We thank Jonah Herzog-Arbeitman, Steven Girvin, Vladimir Calvera, Kyung-Su Kim, Haruki Watanabe, Biao Lian, and Eduardo Fradkin for helpful communications. This work was supported, in part, by the U. S. Department of Energy, Office of Science, Basic Energy Sciences, Materials Sciences and Engineering Division, under Award No. DE-SC0020045.

\bibliographystyle{apsrev4-1} 
\bibliography{ref}

\appendix 

\section{Generalized Witten's conjugation method}
\label{app: conjugation}

Here, we formally discuss the construction scheme we adopted in this work, which could be viewed as a generalization of Witten's conjugation method~\cite{10.21468/SciPostPhysCore.4.4.027,WITTEN1982253}.

Suppose we are given a frustration-free Hamiltonian 
\begin{align}
    \hat{H}=\sum_l \hat{\mathcal{O}}_l^\dagger \hat{\mathcal{O}}_l
\end{align}
and its ground state, $|\Psi_0\rangle$, such that $\hat{\mathcal{O}}_l|\Psi_0\rangle =0$. Then one can construct  new frustration-free models of the form
\begin{align}
    \hat{H}=\sum_l \hat{\mathcal{P}}^{-1,\dagger}\hat{\mathcal{O}}_l^\dagger \hat{\mathcal{M}}_l \hat{\mathcal{O}}_l \hat{\mathcal{P}}^{-1}
\end{align}
with ground state $|\Psi\rangle \equiv \hat{\mathcal{P}}|\Psi_0\rangle$, where $\hat{\mathcal{P}}$ is an arbitrary invertible operator and $\{\hat{\mathcal{M}}_l\}$ is a set of arbitrary positive semi-definite operators. This is because each term in $\hat{H}$ is still positive semi-definite and annihilates $|\Psi\rangle$.

The specific class of models discussed in the main text is of this form with $\hat{\mathcal{O}}_l \rightarrow \hat{B}_{c} -\hat{B}_{c'}$, $\hat{\mathcal{P}} \rightarrow \hat{P}$, and $\hat{\mathcal{M}}_l \rightarrow \hat{P}_c \hat{P}_{c'} \hat{P}^2 $, the later operator so chosen as to ensure that the resulting hopping terms retains a simple and physical form. It is also similar to a construction for  interacting boson models in the continuum studied in Ref.~\cite{PhysRevB.43.3255} where the engineered Hamiltonian is written as
\begin{align}
    \hat{\mathcal{H}} = & \hat{P}^{-1} \left( \nabla \hat{\psi}^\dagger \right)  \hat{P}\cdot \hat{P}\left(\nabla \hat{\psi}\right)\hat{P}^{-1}.
\end{align}
with a similar $\hat{P}$ related to the Boltzmann weight of a classical gas problem in the continuum.

\section{The number basis}
\label{app: number basis}

Since all the $|\alpha\rangle$ have zero energy, any linear combination of them is also a ground state of the system. Since the models are number-conserving, we can obtain a number basis for the ground state manifold:
\begin{align}
    \left|M \right\rangle \equiv & \frac{1}{\mathcal{N}_M}\ \hat{P} \left(\sum_{i}\hat{b}^\dagger_i\right)^M \left|0\right\rangle \\
    =& \frac{M!\mathcal{N}_{|\alpha_0|}}{|\alpha|^M \mathcal{N}_M} \int_{0}^{2\pi} \frac{\mathrm{d}\theta}{2\pi} \mathrm{e}^{-\mathrm{i}M\theta} \left|\alpha=|\alpha_0|\mathrm{e}^{\mathrm{i}\theta}\right\rangle \label{eq: coherent to number basis}
    \ . 
\end{align}
which is true for any fixed non-zero value of $|\alpha_0|$.

As long as $\tau_{c,c'}$ is generic enough for the kinetic Hamiltonian to connect all the possible configuration states $|\{n\}\rangle$ within the corresponding number sector, the ground state $|M\rangle$ is unique (up to potential degeneracies associated with spontaneously broken symmetries or topological order) due to the Perron-Frobenius theorem, since the off-diagonal (kinetic) part of the Hamiltonian in this basis is strictly non-positive and irreducible.

$\mathcal{N}_M$ can be evaluated as:
\begin{align}
    \mathcal{N}_M^2 
    = &\sum_{\{n\}|\sum_k n_k = M} {M \choose n_1, \dots, n_\mathsf{V}}^2 \mathrm{e}^{-K(\{n\})}  \nonumber\\
    &\ \ \ \ \ \ \ \ \ \ \ \ \ \ \ \ \ \ \ \ \ \ \ \langle 0| \prod_k(\hat{b}_k)^{n_k}  (\hat{b}_k^\dagger)^{n_k} |0\rangle \\
    = & (M!)^2\cdot \sum_{\{n\}|\sum_k n_k = M} \frac{1}{\prod_k (n_k!)} \mathrm{e}^{-K(\{n\})} \\
    \equiv & (M!)^2\mathcal{Z}_M \label{eq: canonical mapping}
\end{align}
where $\mathcal{Z}_M$ coincides with the partition function of a classical lattice gas problem in its canonical ensemble. The problem is defined by putting $M$ indistinguishable particles on the lattice sites, assigning each occupation configuration $\{n\}$ an interaction `energy' $K(\{n\})$, and lastly, setting the effective `temperature' of the classical gas to $1$.

\section{The mapping to classical problem}
\label{app: mapping to classical}

Since 
\begin{align}
    \left|\alpha \right\rangle 
    =& \frac{1}{\mathcal{N}_\alpha} \hat{P} \sum_{M=0}^\infty \frac{\alpha^M}{M!} \left(\sum_{i}\hat{b}^\dagger_i\right)^M\left|0\right\rangle \\
    =& \frac{1}{\mathcal{N}_\alpha} \sum_{M=0}^\infty \frac{\alpha^M \mathcal{N}_M |M\rangle }{M!}
\end{align}
$\mathcal{N}_\alpha$ can be evaluated as:
\begin{align}
    \mathcal{N}^2_\alpha =& \sum_M \frac{|\alpha|^{2M}}{(M!)^2} \mathcal{N}^2_M \\
    =& \sum_{\{n\}} \frac{1}{\prod_k (n_k!)} \mathrm{e}^{-K(\{n\}) + \mu_\text{cl} \sum_{k} n_k} \\
    \equiv & \mathcal{Z}_\alpha
\end{align}
where the summation over occupation configurations $\{n\}$ is now unrestricted, and $\mathcal{Z}_\alpha$ can be understood as the partition function of the above classical lattice gas problem in the grand canonical ensemble with $\mu_\text{cl} \equiv \ln(|\alpha|^2)$ being the chemical potential. The total particle number expectation can be thus evaluated as
\begin{align}
   \langle \alpha | \sum_i \hat{n}_i |\alpha \rangle = \frac{\partial \ln \mathcal{Z}_\alpha}{\partial \mu_\text{cl}}
\end{align}

It is straightforward to show that all density correlations in the quantum ground state agree with those in the classical system. We now further show that the boson correlation functions can also be conveniently calculated as with density correlators in the classical system (here $c$ and $c'$ can be of different sizes, $L$ and $L'$):
\begin{widetext}
\begin{align}
        \left\langle \alpha \right| \hat{B}^\dagger_{i_1\dots i_{L_+}} \hat{B}_{j_1\dots j_{L_-}  }   \left|\alpha \right\rangle = & \sum_{\{n\}}   \langle \alpha |\hat{B}^\dagger_{i_1\dots i_{L_+}}  |\{n\}\rangle \langle \{n\} | \hat{B}_{j_1\dots j_{L_-}  } | \alpha \rangle \\
    = & \frac{(\alpha^*)^{L_+}\alpha^{L_-}}{\mathcal{Z}_\alpha} \sum_{\{n\}} \frac{1}{\prod_k (n_k!)}  \mathrm{e}^{-\delta K_{c}(\{n\})/2 - \delta K_{c'}(\{n\})/2} \mathrm{e}^{-K(\{n\}) }  \\
    = & (\alpha^*)^{L_+}\alpha^{L_-}\clbra \mathrm{e}^{-\delta K_{c}(\{n\})/2} \mathrm{e}^{ - \delta K_{c'}(\{n\})/2}\clket_{\alpha}
\end{align}
    \begin{align}
    \left\langle M\right| \hat{B}^\dagger_{c} \hat{B}_{c'} \left|M\right\rangle =& \delta_{L,L'} \sum_{\{n\}|\sum_k n_k = M-L} \left\langle M\right| \hat{B}^\dagger_{c} |\{n\}\rangle \langle \{n\} |\hat{B}_{c'} \left|M\right\rangle \\
    = & \frac{\delta_{L,L'}}{\mathcal{N}^2_M} \sum_{\{n\}|\sum_k n_k = M-L} \frac{1}{\prod_k (n_k!)}  \langle 0| \left(\sum_{i}\hat{b}_i\right)^{M} \hat{P}\hat{B}^\dagger_{c}\prod_k (\hat{b}^\dagger_k)^{n_k}|0\rangle \langle 0 | \prod_k (\hat{b}_k)^{n_k} \hat{B}_{c'} \hat{P} \left(\sum_{i}\hat{b}^\dagger_i\right)^{M}| 0 \rangle  \\
    =& \frac{\delta_{L,L'}}{\mathcal{N}^2_M} \sum_{\{n\}|\sum_k n_k = M-L} \frac{(M!)^2}{\prod_k (n_k!)}  \mathrm{e}^{-\delta K_{c}(\{n\})/2 - \delta K_{c'}(\{n\})/2} \mathrm{e}^{-K(\{n\}) } \\
    = &  \frac{\delta_{L,L'}\mathcal{Z}_{M-L}}{\mathcal{Z}_M}  \clbra \mathrm{e}^{-\delta K_{c}(\{n\})/2} \mathrm{e}^{ - \delta K_{c'}(\{n\})/2} \clket_{M-L}
\end{align}

\end{widetext}

\section{Single mode approximation}
\label{app: dynamical density modes}

The variational energy 
\begin{align}
    E_\text{SMA}(\bm{q})  = \frac{\left\langle \alpha\left| \hat{\rho}_{\bm{q}}^\dagger \hat{H}  \hat{\rho}_{\bm{q}}    \right|\alpha\right\rangle }{\left\langle \alpha\left| \hat{\rho}_{\bm{q}}^\dagger \hat{\rho}_{\bm{q}}    \right|\alpha\right\rangle}
\end{align}
By definition, the denominator is $s(\bm{q})\mathsf{V}$. Then, using $\hat{H} |\alpha\rangle =0$, we can rewrite the numerator as
\begin{align}
\mathsf{V} f(\bm{q}) = \left\langle \alpha\left| \hat{\rho}_{\bm{q}}^\dagger[\hat{H},  \hat{\rho}_{\bm{q}}]    \right|\alpha\right\rangle = \left\langle \alpha\left| \hat{\rho}_{\bm{q}}^\dagger\partial_\tau  \hat{\rho}_{\bm{q}}(\tau)|_{\tau=0}    \right|\alpha\right\rangle 
\end{align}
which is proportional to the time derivative of the dynamical structure factor at equal time and hence its first frequency moment. Due to inversion symmetry we assumed, $f(\bm{q}) = f(-\bm{q}) = f^*(\bm{q})$, so
\begin{align}
    f(\bm{q}) = & \frac{f(\bm{q}) + f(-\bm{q})}{2}\\
    = & \frac{1}{2\mathsf{V}} \left\langle \alpha\left| [\hat{\rho}_{\bm{q}}^\dagger,[\hat{H},  \hat{\rho}_{\bm{q}}] ]  \right|\alpha\right\rangle
\end{align}
Evaluating the double commutator gives rise to the expression in Eq.~\ref{eq: fofq}.

We note that the tightness of the upper bound obtained this way can be further estimated by the variance
\begin{align}
    \left\langle \left(\hat{H}-E_\text{SMA}(\bm{q})\right)^2 \right \rangle = \frac{\left\langle M\right| [\hat{\rho}_{\bm{q}}^\dagger,\hat{H}][\hat{H},  \hat{\rho}_{\bm{q}}]    \left|M\right\rangle}{\left\langle M\right| \hat{\rho}_{\bm{q}}^\dagger \hat{\rho}_{\bm{q}}    \left|M\right\rangle}  - E^2_\text{SMA}(\bm{q})
\end{align}

\section{Generalization to number non-conserving systems}

\label{app: number non conserving}

One can generalize the current construction scheme to number non-conserving systems in several ways. The first is to allow the $c,c'$ in Eq.~\ref{eq: semidefinite} to have different sizes, and modify $\left(\hat{B}_c -\hat{B}_{c'} \right)$ to
\begin{align}\label{eq: number non conserving construction}
    \left(\hat{B}_c/\alpha_0^{L_c} -\hat{B}_{c'}/\alpha_0^{L_{c'}} \right)
\end{align}
where $L_c$ is the size of the cluster $c$. The $\hat{H}_{c,c'}$ constructed in this way does not conserve particle number, and only certain modified coherent states $|\alpha =\alpha_0\mathrm{e}^{\mathrm{i}2\pi r/q}\rangle$ can be shown to be the ground state, where $q = \text{gcd}(L_c,L_{c'})$ and $r = 0,1,\dots q-1$.

A further generalization could be made by defining modified coherent states with a set of site-dependent $\{\alpha_i\}$:
\begin{align}\label{eq: coherent state}
    \left|\{\alpha_i\} \right\rangle \equiv & \frac{1}{\mathcal{N}_{\{\alpha_i\}}}\ \hat{P} \exp\left[ \sum_{i}\alpha_i \hat{b}^\dagger_i \right] \left|0\right\rangle 
    \ . 
\end{align}
then further modifying $\left(\hat{B}_c -\hat{B}_{c'} \right)$ in Eq.~\ref{eq: semidefinite} to 
\begin{align}
    \left[\hat{B}_c/(\prod_{j\in c} \alpha_{0,j}) - \hat{B}_{c'} /(\prod_{j\in c'} \alpha_{0,j})\right]
\end{align}
will give rise to the parent Hamiltonians for ground states $ \left|\{\alpha_i = \alpha_{0,i} \mathrm{e}^{\mathrm{i}2\pi r/q}\} \right\rangle $.

\end{document}